\documentclass[english,aps,prper,reprint,showpacs,longbibliography, floatfix]{revtex4-2}
\usepackage[utf8]{inputenc}
\usepackage{booktabs}
\usepackage{tabularx}
\usepackage{ragged2e}
\usepackage[T1]{fontenc}
\usepackage{geometry}
\usepackage{times}
\usepackage{hyperref}
\hypersetup{colorlinks=true,urlcolor=blue,citecolor=blue,linkcolor=blue}
\usepackage{array}
\usepackage{enumerate}
\usepackage{amsmath}
\usepackage{amssymb}
\usepackage{tikz}
\usepackage{graphicx}
\usepackage{multirow}
\usepackage{tcolorbox}
\usepackage{float}
\usepackage[normalem]{ulem}
\usepackage{setspace}
\usepackage{epstopdf}

\begin{document}

\title{Analyzing Undergraduate Problem-Solving in Physics Through  Interaction With an AI Chatbot}

\author{Syed Furqan Abbas Hashmi}
\affiliation{Department of Physics and Astronomy, Purdue University, West Lafayette, IN 47907, U.S.A.}
\author{N. Sanjay Rebello}
\affiliation{Department of Physics and Astronomy \& Department of Curriculum \& Instruction, Purdue University, West Lafayette, IN 47907, U.S.A.}

 \begin{abstract}
Providing individualized scaffolding for physics problem solving at scale remains an instructional challenge. We investigate (1) students’ perceptions of a Socratic Artificial Intellegence (AI) chatbot’s impact on problem‐solving skills and confidence and (2) how the specificity of students’ questions during tutoring relates to performance. We deployed a custom Socratic AI chatbot in a large‐enrollment introductory mechanics course at a Midwestern public university, logging full dialogue transcripts from 150 first‐year STEM majors. Post‐interaction surveys revealed median ratings of 4.0/5 for knowledge‐based skills and 3.4/5 for overall effectiveness. Transcript analysis showed question specificity rose from approximately 10–15\% in the first turn to 100 \% by the final turn, and specificity correlated positively with self-reported expected course grade (Pearson $r\approx0.43$). These findings demonstrate that AI-driven Socratic dialogue not only fosters expert-like reasoning but also generates fine-grained analytics for physics education research, establishing a scalable dual-purpose tool for instruction and learning analytics.

\end{abstract}

\maketitle

\section{Introduction}

Problem solving is widely recognized as the fundamental means by which students demonstrate conceptual mastery and quantitative reasoning in physics \cite{Reif1982, PhysRevSTPER.10.020119, PhysRevPhysEducRes.13.020101}.  Yet, despite decades of discipline‐based research revealing expert–novice differences in strategic planning, representation use, and metacognitive monitoring \cite{Chi1981, PhysRevSTPER.4.010111, PhysRevSTPER.3.020101}, traditional assessments continue to capture only final answers and sparse work artifacts, rendering the rich cognitive processes invisible \cite{PhysRevPhysEducRes.12.010130, PhysRevPhysEducRes.18.010109}.  This invisibility presents a critical gap: without real‐time data on how students engage in problem solving, instructors cannot scaffold expert‐like strategies nor provide targeted feedback.

Recent advances in Artificial Intellegence (AI) and Natural Language Processing (NLP) offer a powerful solution.  Intelligent tutoring systems and AI chatbots have demonstrated near human‐level gains in one‐on‐one settings, often approaching the “two‐sigma” benchmark of human tutoring effectiveness \cite{VanLehn_2011, Graesser2005, PhysRevPhysEducRes.21.010153}.  Generative‐AI tools have been applied to grade explanations and generate feedback with human‐comparable accuracy \cite{PhysRevPhysEducRes.21.010126, PhysRevPhysEducRes.21.010128}, and data‐augmentation studies have shown how synthetic responses can train reliable analytic models \cite{PhysRevPhysEducRes.19.020150}.  However, these innovations primarily focus on aggregate performance metrics or educator training applications \cite{PhysRevPhysEducRes.21.010154, PhysRevPhysEducRes.20.010152, PhysRevPhysEducRes.20.020144, PhysRevPhysEducRes.20.010109, PhysRevPhysEducRes.19.020163}, leaving a second gap: few studies integrate Socratic dialogue systems that both scaffold metacognitive reflection and capture complete interaction transcripts for fine‐grained analytics.

To address these gaps, we developed and deployed a custom Socratic AI physics chatbot in an introductory mechanics course at a large U.S. Midwestern university.  Our system prompts students with targeted, sequential questions, logs full dialogue transcripts, and applies automated analytics to quantify question specificity and problem‐solving progress in real time \cite{SIRNOORKAR2024100318}.  We investigate two primary research questions:
\begin{itemize}
  \item How do students perceive the chatbot’s impact on their problem‐solving skills and confidence?
  \item How does the specificity of students’ questions during the tutoring dialogue relate to their problem‐solving progress and performance?
\end{itemize}
By combining AI‐driven scaffolding with research‐grade data streams, this work advances both classroom pedagogy and PER methodology.

\section{Theoretical Bases}

Our design builds on established frameworks for expert problem solving and Socratic inquiry.  Adams and Wieman’s multi‐skill model identifies conceptual understanding, strategic planning, and self‐monitoring as core sub‐skills in physics problem solving \cite{Adams_Wieman_2015, Chi_Bassok_Lewis_Reimann_Glaser_1989}.  Epistemic‐game theory further maps the cognitive moves experts use to structure solutions \cite{PhysRevSTPER.3.020101}, while multiple‐representation studies highlight the importance of fluid transitions among algebraic, graphical, and pictorial modes \cite{PhysRevSTPER.4.010111}.  Rubric‐based assessments provide reliable snapshots of written work \cite{PhysRevPhysEducRes.12.010130, PhysRevPhysEducRes.13.020101}, and metacognitive post‐reflection exercises yield measurable gains in strategic reasoning \cite{PhysRevPhysEducRes.18.010109, PhysRevPhysEducRes.13.020101}.

Socratic tutoring, which uses targeted questions to elicit self‐explanations, has long been shown to enhance learning by revealing and correcting misconceptions \cite{Graesser2005, Chi_Bassok_Lewis_Reimann_Glaser_1989}.  Intelligent tutoring systems that implement Socratic dialogue achieve substantial learning gains in physics and other domains \cite{VanLehn_2011}.  More recently, PER has begun to explore AI chatbots: Dunlap et al.\ demonstrated Large Language Model (LLM) performance on kinematics graphs \cite{PhysRevPhysEducRes.20.010109}, and Polverini et al.\ evaluated ChatGPT on the BEMA inventory and concluded that ChatGPT exhibits significant difficulties in engaging with physics tasks involving visual representations. \cite{PhysRevPhysEducRes.21.010154}.  Fussell et al.\ compared LLMs for lab‐note analysis \cite{PhysRevPhysEducRes.21.010128}, and Chen \& Wan showed human‐level feedback generation on problem‐solving explanations \cite{PhysRevPhysEducRes.21.010126}.  Kieser et al.\ applied data augmentation with ChatGPT to train analytic models \cite{PhysRevPhysEducRes.19.020150}, and Kortemeyer et al.\ explored AI grading feasibility on handwritten exams \cite{PhysRevPhysEducRes.20.020144, PhysRevPhysEducRes.19.020163}.

Despite these advances, transcript‐based dialogue analytics remain scarce.  Existing studies tend to focus on performance metrics or teacher training, rather than authentic student–bot interactions parsed for question specificity and linked directly to learning outcomes \cite{PhysRevPhysEducRes.21.010153, SIRNOORKAR2024100318}.  By integrating Socratic scaffolding with automated analytics on complete dialogue transcripts, our work fills this critical methodological and pedagogical gap in PER.

\section{Methods}
The study took place in an first-semester, calculus-based physics course at a large U.S. Midwestern public university. A total of $N = 150$ first-year STEM majors participated in the study which was conducted in Week 9 of the Spring 2025 semester. At the start of each session, students received a brief orientation to the Socratic chatbot and were instructed: “Interact with this tool just as you would with a human teaching assistant.” Students then individually engaged with the chatbot via a web interface (a Django web application deployed on Heroku) using their own laptops or phones. The chatbot presented a multi-step physics problem scenario and guided the student through it in dialogue. The problem shown in Figure ~\ref{fig:problem statement}  (adapted from a Context Rich Problem \cite{Heller_Keith_Anderson_1992} from the University of Minnesota PER group website) involved a human cannonball being launched by a spring -- requiring energy conservation, projectile motion, and kinematics concepts to determine if the performer would clear a 15-foot wall. This rich context required students to integrate Hooke’s law, gravitational potential energy, and projectile motion principles. The chatbot was pre-loaded with the problem statement (which students could review at any time) and began by asking the student to consider the given data and “What is your approach?” or similar broad prompts.
\begin{figure}[!t]
    \centering
    \includegraphics[width=0.8\columnwidth]{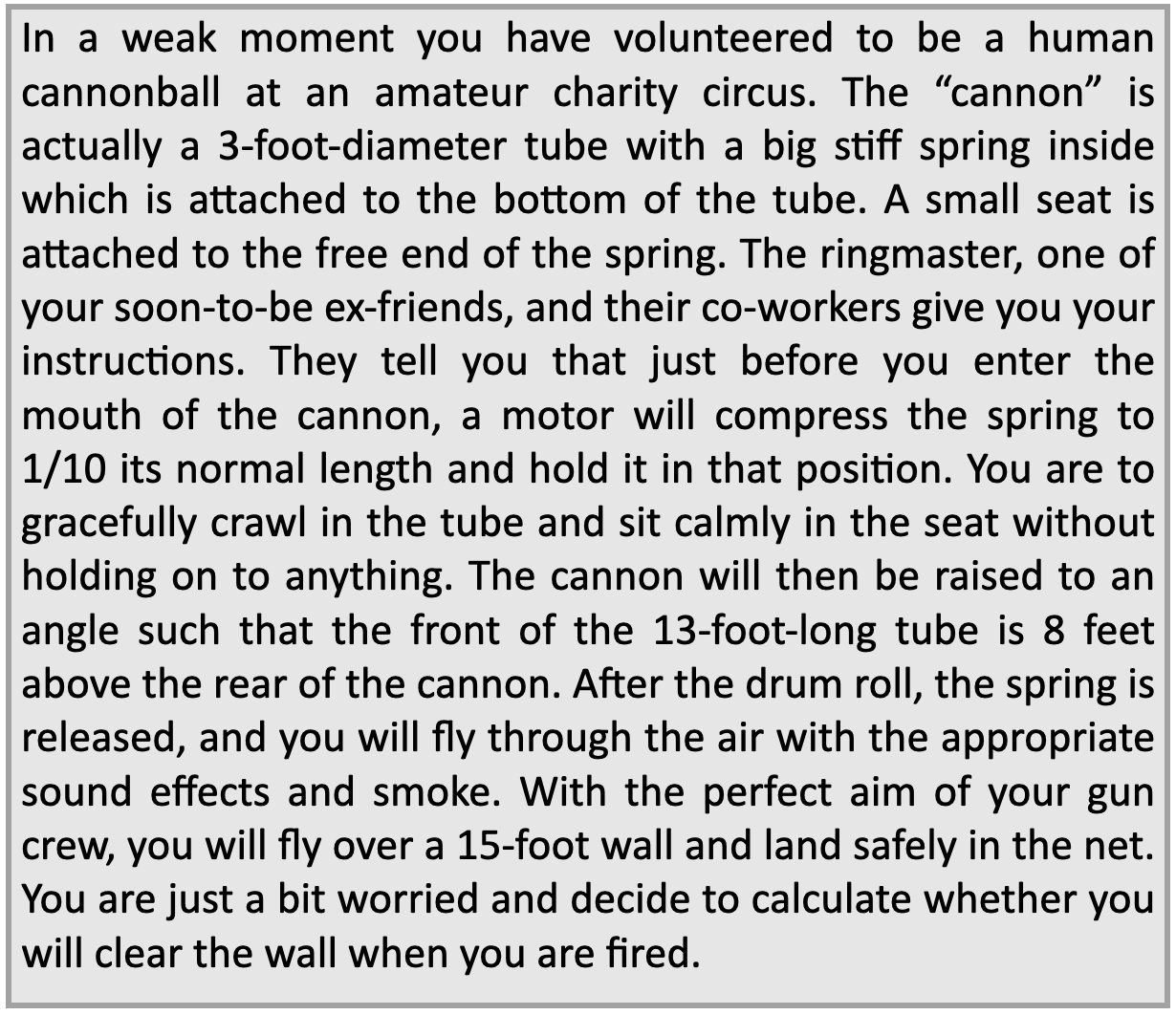}
    \caption{Problem statement of the context rich problem the students were asked to solve using the chatbot.}
    \label{fig:problem-statement}
\end{figure}
The AI tutor’s system architecture was designed to emulate an interactive Socratic dialogue while ensuring correctness of the solution path. The backend (Django) maintained session state and managed Application Programming Interface (API) calls to OpenAI’s GPT-4o model. Two GPT-4o instances were used: one as the Instructor which interacted with the stedent and responded to them, and one as the Verifier, which would verify the correctness of the student response. The Instructor is responsible for generating the next question or hint for the student, whereas the Verifier evaluates the student’s latest answer by using ChatGPT's default knowledge-base. The problem-solving dialogue is structured as a sequence of steps (e.g., conceptual understanding, setting up equations, performing calculations). At each step, the Instructor poses a question. When the student submits an answer, the Verifier agent is invoked to check the answer’s correctness and quality.

All student–chatbot conversations were then submitted by the students. After the 50-minute session (or when a student finished earlier), students were directed to complete an online survey about their experience. The survey was administered via Qualtrics and consisted of 20 Likert-scale items (5-point agreement scale) targeting different dimensions of problem-solving. These items were developed with reference to Adams and Wieman’s problem-solving framework \cite{Adams_Wieman_2015}. The survey covered: (1) Knowledge -- Based Skills (e.g. understanding concepts, applying prior knowledge), (2) Beliefs, Expectations \& Motivation (e.g. confidence after using the chatbot, enjoyment, willingness to persist), (3) Problem -- Solving Processes (e.g. planning, checking work, connecting concepts, identifying irrelevant information), (4) Task Difficulty Comparison (e.g. how difficult was it to interact with the chatbot), and (5) Overall Effectiveness (e.g. whether the chatbot helped solve problems independently, improved approach to new problems, and if the student would recommend it). Students’ expected course grade was also collected as a single self-reported item at the end of the survey. To ensure data quality, only fully completed surveys were included in analysis. In total, we received 150 complete responses.

Each item asked students to rate their agreement with statements about the chatbot’s impact on their problem-solving experience during the session. We note that the survey, being administered immediately post-activity, captures students’ perceptions of their learning and engagement rather than direct learning outcomes. Nonetheless, these self-reports provide insight into which aspects of the problem-solving process the students felt were enhanced by the chatbot.

To investigate students’ help‐seeking behavior, we analyzed each conversation transcript for the types of questions students asked the chatbot. We were particularly interested in \emph{question specificity}—whether a student’s query was broad/general or specific to a physics concept or step. We operationally defined a \emph{broad question} as one that expressed general confusion or a very open‐ended request (e.g., “I don’t know how to start” or “What should I do next?”). In contrast, a \emph{specific question} referenced a particular concept, principle, or calculation related to the problem (e.g., “Should I use conservation of energy here?”, or “Do I need to account for the initial height in the spring energy?”). Each student utterance that was a question to the chatbot was coded as \emph{broad} or \emph{specific} based on this criterion. Most student–tutor dialogues consisted of about 5–6 student question turns on average (since the tutor guided them through roughly that many sub‐problems). For each student, we calculated the proportion of their questions that were \emph{specific} (versus \emph{broad}) across the session. We also noted how this specificity changed from the beginning to the end of the dialogue (i.e., whether there was a shift toward more specific questions over time). Finally, we examined the relationship between question‐asking patterns and the student’s expected grade in the course, using expected grade as a metric for overall performance.

\section{Results}
\subsection{Survey Results}

\begin{figure}[ht]
  \centering
  \includegraphics[width=\columnwidth]{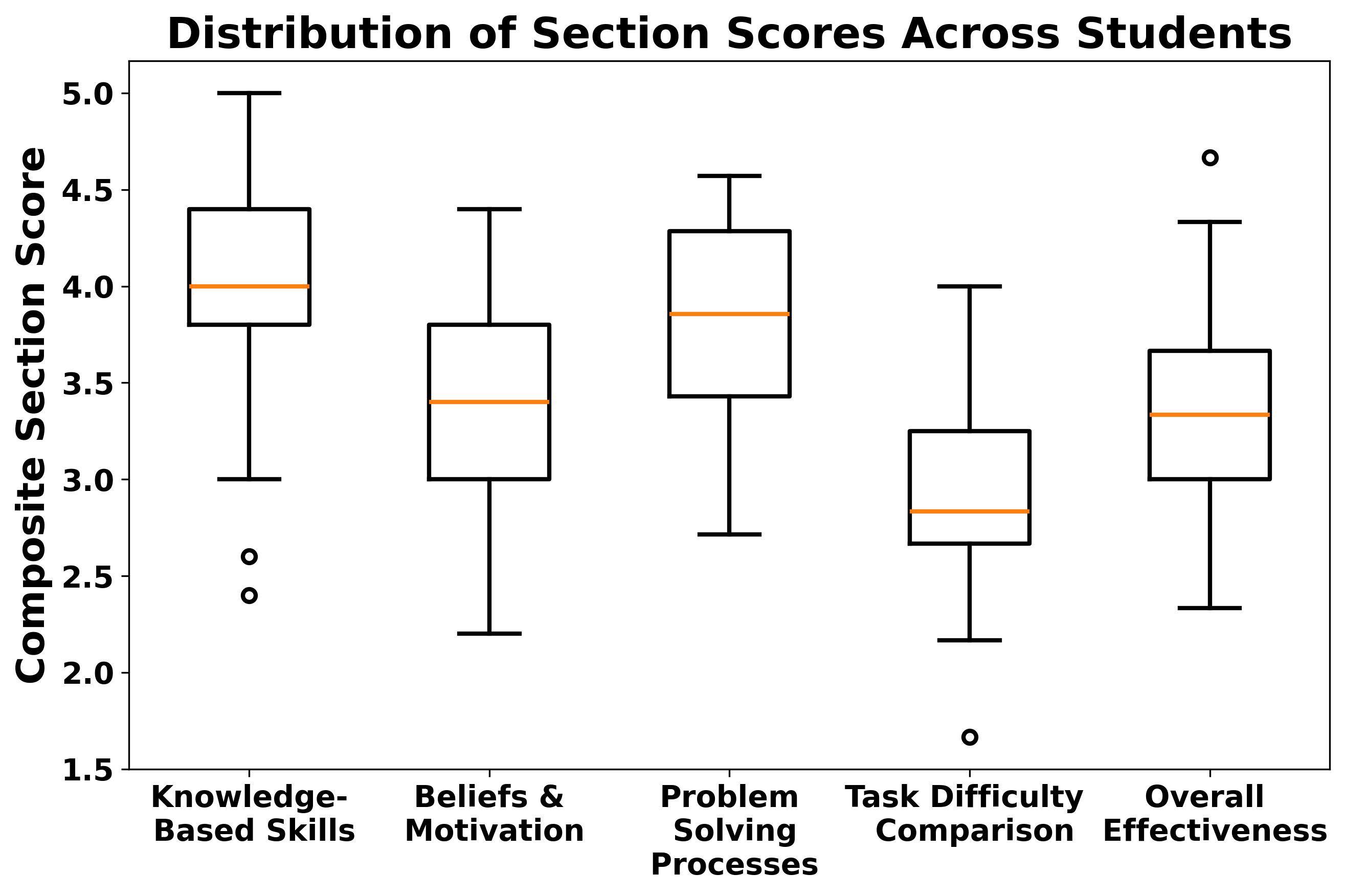}
  \caption{Distribution of composite survey scores for the five dimensions of the chatbot impact.}
  \label{fig:survey_dimensions}
\end{figure}

Student responses to the post‐activity survey were generally positive, indicating that the AI tutor was effective in supporting several problem‐solving skills. To get an overview of all survey dimensions, Figure~\ref{fig:survey_dimensions} presents the distribution of composite scores for each section of the survey. We computed an average rating for the items in a given section (Knowledge, Motivation, Problem -- Solving Skills, Difficulty, Overall Effectiveness). The box plots in Figure~\ref{fig:survey_dimensions} represent these results. We observe that Knowledge‐Based Skills had the highest overall ratings (median $\approx 4.0$). The Overall Effectiveness section was also rated positively (median $\approx 3.4$), indicating that many students would recommend the chatbot and felt it improved their independent problem‐solving ability. Problem‐Solving Process section had a median $\approx 3.9$ and a more narrow spread. Some students reported that the chatbot greatly improved their planning and checking strategies, others were more neutral on these process‐oriented items. The Beliefs/Motivation dimension in Figure~\ref{fig:survey_dimensions}) showed moderate ratings (median $\approx 3.4$), suggesting that affective gains (confidence, enjoyment, perseverance) were present but not as universally high as the cognitive gains. Notably, the Difficulty dimension refers to students’ perception of problem difficulty with the chatbot’s help (these items asked whether the problems felt easier or more manageable). This survey highlights that the chatbot did not entirely eliminate the challenge of the task, especially for those who remained confused. Overall, the survey data suggest that the AI tutor was most effective in helping students learn physics content and problem‐solving methods, while providing a moderate boost to confidence and motivation.

\subsection{Dialogue Patterns and Question Specificity}

Analysis of the chatbot transcripts provided insight into how students’ questioning behavior evolved during the tutoring sessions. Overall, we found a general progression from broad questions to specific questions as the dialogue advanced. At the very beginning of the interaction, most students asked rather broad questions or expressed uncertainty. For example, it was common for a student’s first query to be something like “I don’t know how to start this” or “What should I do first?”. In the first turn, only about 10–15\% of students asked a specific physics question. However, as the session continued, students’ questions became more targeted. By the fourth turn in the dialogue, approximately 58\% of students were asking specific, content‐focused questions, such as referring to a particular principle or calculation needed. By the final turn of the tutoring conversation, all of students were asking specific questions rather than broad ones.

\begin{figure}[ht]
  \centering
  \includegraphics[width=\columnwidth]{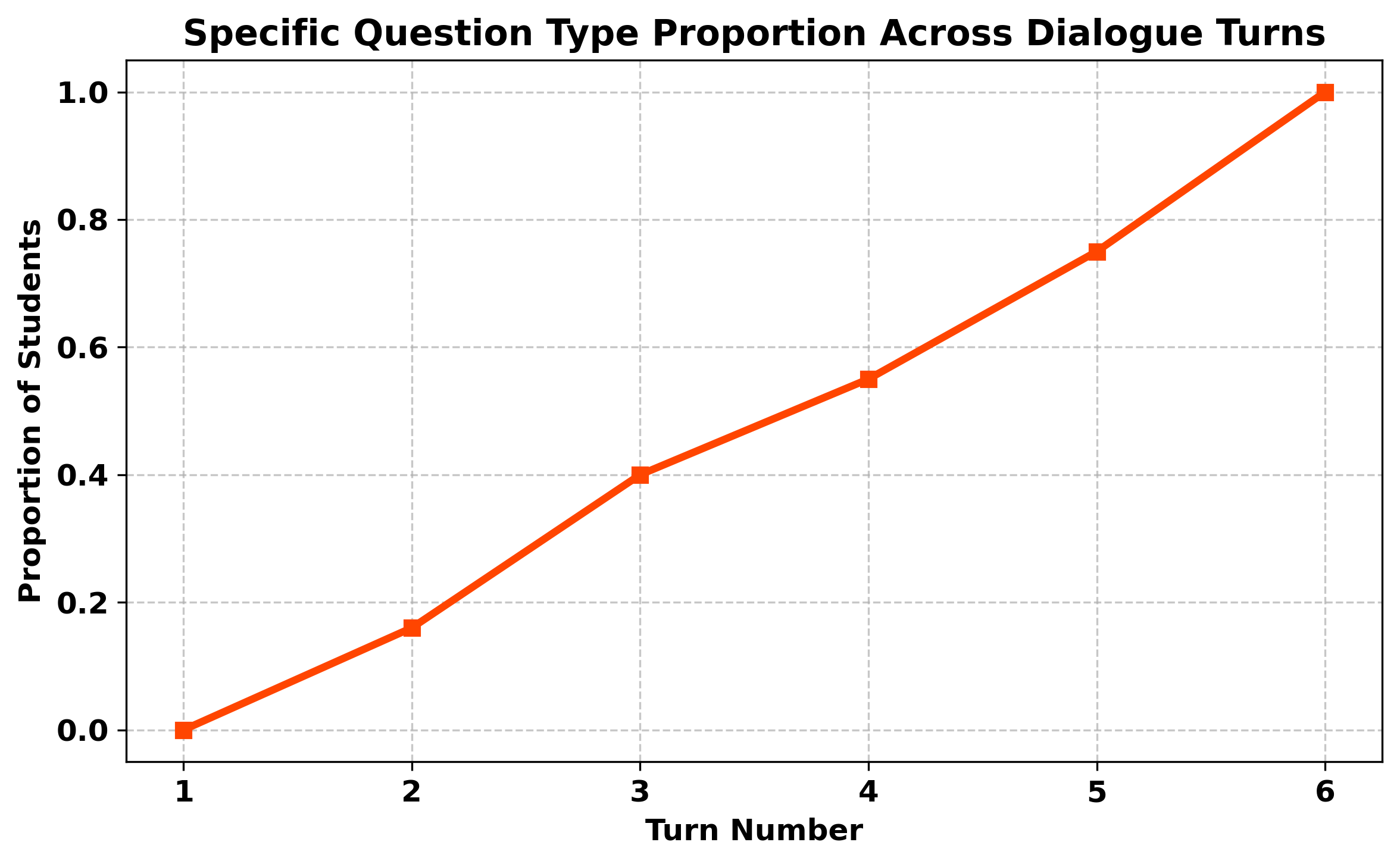}
  \caption{Proportion of student questions classified as specific at each turn of the dialogue.}
  \label{fig:question_specificity}
\end{figure}

\subsection{Question Specificity and Expected Grade}

\begin{figure}[ht]
  \centering
  \includegraphics[width=\columnwidth]{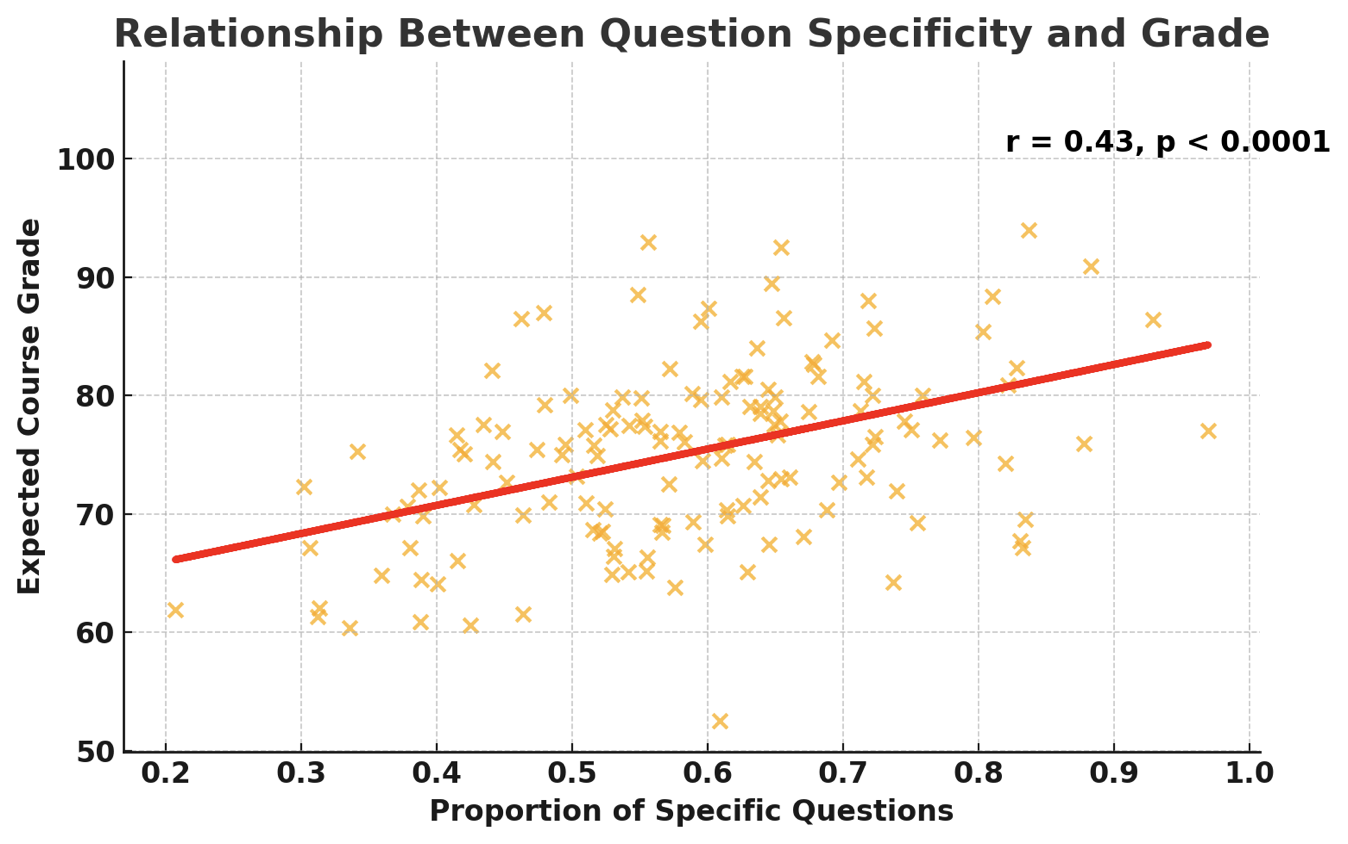}
  \caption{Relationship between question specificity and expected course grade. Each point represents one student, with the $x$‐axis showing the fraction of their questions that were specific and the $y$‐axis showing their expected grade. The orange line is a linear fit (Pearson $r \approx 0.43$).}
  \label{fig:spec_vs_grade}
\end{figure}

We next examined whether students who asked more specific questions tended to perform better, as indicated by their self‐reported expected course grade, based on their performance in Week 9 of the semester. Figure~\ref{fig:spec_vs_grade} shows the relationship between each student’s question specificity (measured as the fraction of their questions that were specific) and their expected grade in the course. A mild positive correlation was observed (Pearson $r \approx 0.43,\hphantom{x} p < 0.0001$). In general, students who engaged in a higher proportion of specific, detailed questioning during the chatbot session reported higher expected grades. For instance, several students who expected to earn an “A” had over 80\% of their questions classified as specific. Conversely, some students with mostly broad questions tended to report lower expected grades (around the “C” range). This pattern aligns with our hypothesis that more adept problem solvers would utilize the chatbot in a more focused way—drilling into particular concepts—whereas less prepared students might remain stuck asking very general questions.

However, the relationship was far from perfectly predictive. As seen in Figure~\ref{fig:spec_vs_grade}, there is considerable scatter in the data. A few students with high specificity (near 1.0) still only anticipated a mid‐range grade, and some students who expected high grades asked surprisingly broad questions throughout the session. The linear fit line in Figure~\ref{fig:spec_vs_grade} illustrates the positive trend but also highlights the large residuals. Thus, while asking more specific questions appears to be associated with somewhat better performance, it is not the sole determinant of success. Other factors—such as prior knowledge, general academic ability, or degree of engagement with the chatbot -- likely also play significant roles.

It is worth noting that we also explored connections between the survey results and the dialogue features. For instance, we expected that students who rated the chatbot highly in Knowledge or Problem-Solving Process (survey sections 1 and 3) might also have engaged in more specific questioning and performed well. Indeed, we observed some directional consistency with this expectation: students with higher Knowledge -- Based survey scores often had above-average question specificity and higher grades. Those who reported lower confidence or motivation (survey section 2) tended to ask more broad questions and sometimes struggled more. 
In general, the data paint a coherent picture: students who fully embraced the guided problem-solving (asking targeted questions, using the tutor’s feedback) also felt they learned more and in some cases achieved better outcomes. Meanwhile, students who remained lost (asking broad questions even late in the session) were more likely to give lukewarm survey feedback and have lower performance. These insights, though preliminary, suggest that the richness of student -- AI dialogues can potentially serve as a window into students’ problem-solving behaviors and needs.

\section{Conclusions}
In response to our first research question: \emph{How do students perceive the chatbot’s impact on their problem -- solving skills and confidence?}  The post-activity survey (Figure~\ref{fig:survey_dimensions}) shows that Socratic AI dialogue enhanced their conceptual understanding and problem solving, 
Turning to our second research question —\emph{How does the specificity of students’ questions relate to their problem-solving progress and performance?} Our analysis of the chatbot transcripts (Figure~\ref{fig:question_specificity}) reveals a clear shift from broad to targeted queries, rising from ~10–15\% specific in Turn 1 to ~75\% by the final turn. Moreover, question specificity correlates mildly but positively with expected course grade (Pearson $r\approx0.43$, Figure~\ref{fig:spec_vs_grade}), showing that more focused inquiry is associated with stronger performance indicators. The scatter around this trend, however, suggests that question specificity is one of multiple factors driving success.

\section{Limitations \& Future Work}
While encouraging, this study has several key limitations. First, it was conducted in a single course using one problem scenario (the human‐cannonball spring launch) with first‐year students, so our findings may not generalize; future work will deploy the chatbot across the length of a full course and problem types. Second, we used a coarse binary broad/specific question classification that overlooks important nuances, some “specific” questions probe conceptual understanding, while some “broad” ones simply verify approach. We will develop a finer‐grained taxonomy (e.g., conceptual, procedural, verification, motivational) and leverage NLP to capture subtleties like uncertainty or misconception. Third, we relied on self‐reported expected grades and survey perceptions rather than direct measures of learning; subsequent studies will include objective post‐tests, control groups, and long‐term retention tracking to assess actual gains.

From a system standpoint, the fixed question progression and GPT--4o hints limit adaptivity. We plan to integrate sequence -- based dialogue models (e.g., transformer classifiers) and multimodal signals (response timing, hint requests) to detect when a student is stuck and adjust difficulty on the fly. As dialogue data accumulate, adaptive feedback loops can identify common misconceptions and refine the Verifier’s scaffolding strategies.

Finally, the chatbot is not meant to replace human instructors or peer collaboration. The uneven motivation gains suggest that AI tutoring should be complemented by group recitations or instructor guidance. Moving forward, we will broaden the scope, deepen dialogue analytics, and enhance pedagogical interventions to create scalable, effective support for STEM problem -- solving.

\section*{Acknowledgements}
This work is supported in part by U.S. National Science Foundation grant 2300645. Opinions expressed are those of the authors and not of the Foundation.

\clearpage
\bibliographystyle{unsrt}
\bibliography{references}

\begin{thebibliography}{10}

\bibitem{Reif1982}
F.~Reif and Joan~I. Heller.
\newblock Knowledge structure and problem solving in physics.
\newblock {\em Educational Psychologist}, 17(2):102--127, 1982.

\bibitem{PhysRevSTPER.10.020119}
Jennifer~L. Docktor and Jos\'e~P. Mestre.
\newblock Synthesis of discipline-based education research in physics.
\newblock {\em Phys. Rev. ST Phys. Educ. Res.}, 10:020119, Sep 2014.

\bibitem{PhysRevPhysEducRes.13.020101}
Anne~E. Leak, Susan~L. Rothwell, Javier Olivera, Benjamin Zwickl, Jarrett Vosburg, and Kelly~Norris Martin.
\newblock Examining problem solving in physics-intensive ph.d. research.
\newblock {\em Phys. Rev. Phys. Educ. Res.}, 13:020101, Jul 2017.

\bibitem{Chi1981}
Michelene T.~H. Chi, Paul~J. Feltovich, and Robert Glaser.
\newblock Categorization and representation of physics problems by experts and novices.
\newblock {\em Cognitive Science}, 5(2):121--152, 1981.

\bibitem{PhysRevSTPER.4.010111}
Patrick~B. Kohl and Noah~D. Finkelstein.
\newblock Patterns of multiple representation use by experts and novices during physics problem solving.
\newblock {\em Phys. Rev. ST Phys. Educ. Res.}, 4:010111, Jun 2008.

\bibitem{PhysRevSTPER.3.020101}
Jonathan Tuminaro and Edward~F. Redish.
\newblock Elements of a cognitive model of physics problem solving: Epistemic games.
\newblock {\em Phys. Rev. ST Phys. Educ. Res.}, 3:020101, Jul 2007.

\bibitem{PhysRevPhysEducRes.12.010130}
Jennifer~L. Docktor, Jay Dornfeld, Evan Frodermann, Kenneth Heller, Leonardo Hsu, Koblar~Alan Jackson, Andrew Mason, Qing~X. Ryan, and Jie Yang.
\newblock Assessing student written problem solutions: A problem-solving rubric with application to introductory physics.
\newblock {\em Phys. Rev. Phys. Educ. Res.}, 12:010130, May 2016.

\bibitem{PhysRevPhysEducRes.18.010109}
Aaron Reinhard, Alex Felleson, Paula~C. Turner, and Maxwell Green.
\newblock Assessing the impact of metacognitive postreflection exercises on problem-solving skillfulness.
\newblock {\em Phys. Rev. Phys. Educ. Res.}, 18:010109, Jan 2022.

\bibitem{VanLehn_2011}
Kurt VanLehn.
\newblock The relative effectiveness of human tutoring, intelligent tutoring systems, and other tutoring systems.
\newblock {\em Educational Psychologist}, 46(4):197--221, 2011.

\bibitem{Graesser2005}
Arthur~C Graesser, Katja Wiemer-Hastings, Peter Wiemer-Hastings, and Roger Kreuz.
\newblock Autotutor: A simulation of a human tutor.
\newblock {\em Cognitive Systems Research}, 1(1):35--51, 1999.

\bibitem{PhysRevPhysEducRes.21.010153}
Justin~C. Dunlap, Ryan Sissons, and Ralf Widenhorn.
\newblock Descending an inclined plane with a large language model.
\newblock {\em Phys. Rev. Phys. Educ. Res.}, 21:010153, May 2025.

\bibitem{PhysRevPhysEducRes.21.010126}
Zhongzhou Chen and Tong Wan.
\newblock Grading explanations of problem-solving process and generating feedback using large language models at human-level accuracy.
\newblock {\em Phys. Rev. Phys. Educ. Res.}, 21:010126, Mar 2025.

\bibitem{PhysRevPhysEducRes.21.010128}
Rebeckah~K. Fussell, Megan Flynn, Anil Damle, Michael F.~J. Fox, and N.~G. Holmes.
\newblock Comparing large language models for supervised analysis of students' lab notes.
\newblock {\em Phys. Rev. Phys. Educ. Res.}, 21:010128, Mar 2025.

\bibitem{PhysRevPhysEducRes.19.020150}
Fabian Kieser, Peter Wulff, Jochen Kuhn, and Stefan Küchemann.
\newblock Educational data augmentation in physics education research using chatgpt.
\newblock {\em Phys. Rev. Phys. Educ. Res.}, 19:020150, Oct 2023.

\bibitem{PhysRevPhysEducRes.21.010154}
Giulia Polverini, Jakob Melin, Elias Önerud, and Bor Gregorcic.
\newblock Performance of chatgpt on tasks involving physics visual representations: The case of the brief electricity and magnetism assessment.
\newblock {\em Phys. Rev. Phys. Educ. Res.}, 21:010154, May 2025.

\bibitem{PhysRevPhysEducRes.20.010152}
Tong Wan and Zhongzhou Chen.
\newblock Exploring generative ai assisted feedback writing for students' written responses to a physics conceptual question with prompt engineering and few-shot learning.
\newblock {\em Phys. Rev. Phys. Educ. Res.}, 20:010152, Jun 2024.

\bibitem{PhysRevPhysEducRes.20.020144}
Gerd Kortemeyer, Julian N\"ohl, and Daria Onishchuk.
\newblock Grading assistance for a handwritten thermodynamics exam using artificial intelligence: An exploratory study.
\newblock {\em Phys. Rev. Phys. Educ. Res.}, 20:020144, Nov 2024.

\bibitem{PhysRevPhysEducRes.20.010109}
Giulia Polverini and Bor Gregorcic.
\newblock Performance of chatgpt on the test of understanding graphs in kinematics.
\newblock {\em Phys. Rev. Phys. Educ. Res.}, 20:010109, Feb 2024.

\bibitem{PhysRevPhysEducRes.19.020163}
Gerd Kortemeyer.
\newblock Toward ai grading of student problem solutions in introductory physics: A feasibility study.
\newblock {\em Phys. Rev. Phys. Educ. Res.}, 19:020163, Nov 2023.

\bibitem{SIRNOORKAR2024100318}
Amogh Sirnoorkar, Dean Zollman, James~T. Laverty, Alejandra~J. Magana, N.~Sanjay Rebello, and Lynn~A. Bryan.
\newblock Student and ai responses to physics problems examined through the lenses of sensemaking and mechanistic reasoning.
\newblock {\em Computers and Education: Artificial Intelligence}, 7:100318, 2024.

\bibitem{Adams_Wieman_2015}
Wendy~K. Adams and Carl~E. Wieman.
\newblock Analyzing the many skills involved in solving complex physics problems.
\newblock {\em American Journal of Physics}, 83(5):459--467, 05 2015.

\bibitem{Chi_Bassok_Lewis_Reimann_Glaser_1989}
Michelene~T.H. Chi, Miriam Bassok, Matthew~W. Lewis, Peter Reimann, and Robert Glaser.
\newblock Self-explanations: How students study and use examples in learning to solve problems.
\newblock {\em Cognitive Science}, 13(2):145--182, 1989.

\bibitem{Heller_Keith_Anderson_1992}
Patricia Heller, Ronald Keith, and Scott Anderson.
\newblock Teaching problem solving through cooperative grouping. part 1: Group versus individual problem solving.
\newblock {\em Am. J. Phys.}, 60(7):627--636, July 1992.

\end{thebibliography}
\end{document}